\documentclass[11pt,preprint]{aastex}
\usepackage{multirow}

\shorttitle{UV photodesorption of methanol in pure and CO-rich ices}
\shortauthors{Bertin et al.}

\begin{document}

\title{UV photodesorption of methanol in pure and CO-rich ices: desorption rates of the intact molecule and of the photofragments }

\author{Mathieu Bertin\altaffilmark{1}, Claire Romanzin\altaffilmark{2}, Mikhail Doronin\altaffilmark{1}, Laurent Philippe\altaffilmark{1}, Pascal Jeseck\altaffilmark{1}, Niels Ligterink\altaffilmark{3}, Harold Linnartz\altaffilmark{3}, Xavier Michaut\altaffilmark{1} and Jean-Hugues Fillion\altaffilmark{1}}

\altaffiltext{1}{LERMA, Sorbonne Universit\'es, UPMC Univ. Paris 06, Observatoire de Paris, PSL Research University, CNRS, F-75252, Paris, France.}
\altaffiltext{2}{LCP (UMR 8000), CNRS, Universit\'e Paris-Sud, 91405 Orsay, France}
\altaffiltext{3}{Sackler Laboratory for Astrophysics, Leiden Observatory, Leiden University, P.O. Box 9513, NL-2300 RA Leiden, The Netherlands.}

\begin{abstract}
Wavelength dependent photodesorption rates have been determined using synchrotron radiation, for condensed pure and mixed methanol ice in the 7 -- 14 eV range. The VUV photodesorption of intact methanol molecules from pure methanol ices is found to be of the order of 10$^{-5}$ molecules/photon, that is two orders of magnitude below what is generally used in astrochemical models. This rate gets even lower ($<$ 10$^{-6}$ molecules/photon) when the methanol is mixed with CO molecules in the ices. This is consistent with a picture in which photodissociation and recombination processes are at the origin of intact methanol desorption from pure CH$_3$OH ices. Such low rates are explained by the fact that the overall photodesorption process is dominated by the desorption of the photofragments CO, CH$_3$, OH, H$_2$CO and CH$_3$O/CH$_2$OH, whose photodesorption rates are given in this study. Our results suggest that the role of the photodesorption as a mechanism to explain the observed gas phase abundances of methanol in cold media is probably overestimated. Nevertheless, the photodesorption of radicals from methanol-rich ices may stand at the origin of  the gas phase presence of radicals such as CH$_3$O, therefore opening new gas phase chemical routes for the formation of complex molecules. 
\end{abstract}


\section{Introduction}

The methanol (CH$_3$OH) molecule is an important species in the chemistry of the interstellar medium (ISM). It is often considered as a starting point for the synthesis of more complex organics, ultimately leading to the formation of prebiotic molecules such as simple amino-acids \citep{mun02,gar08}. Since gas phase chemistry cannot account alone for the observed methanol abundances \citep{gar06,gep06}, it is generally accepted that its formation takes place in the condensed phase, by successive hydrogenation steps in CO-rich icy mantles \citep{wat02,fuc09}. The presence of methanol as a major constituent of the icy grain mantles is indeed confirmed by observations (e.g. \cite{gib00,boo11}), with variable importance for different lines of sight \citep{whi11}. In the gas phase,  CH$_3$OH is observed in hot cores (e.g. \cite{cec00}), where the temperature of the grains gets sufficiently high for the condensed methanol to thermally desorb. However, gas phase methanol is also observed in cold regions (T $\sim$ 10 -- 30 K), such as prestellar cores \citep{vas14} or low-UV-field illumination photo-dissociation regions (PDRs) \citep{guz13}, suggesting that other, non-thermal desorption mechanisms must be at play to explain the gas phase  abundance. These processes can be driven by the impact of high energy particles in the ice, or by surface exothermic recombination reactions referred to as reactive or chemical desorption \citep{vas13, dul13}. Among them, the desorption induced by UV photons, so-called UV photodesorption, is generally invoked to explain the CH$_3$OH gaseous abundances. Astrochemical models which comprise grain-gas interactions need as input absolute photodesorption rates of methanol to reproduce accurately the observational findings \citep{guz13,vas14}. Photodesorption rates are also needed to predict gas phase CH$_3$OH abundances and snowline locations in regions where it has not yet been detected, such as in protoplanetary disks \citep{wal14}. 

A number of laboratory experiments has been dedicated to the study of photodesorption of several simple astrochemically relevant molecules (CO, O$_2$, O$_3$ H$_2$O, N$_2$, CO$_2$...) these last ten years (e.g. \cite{obe07,mun10,fay11,zhe14,yua13,bah12}). In particular, the photodesorption and photochemistry of pure methanol ices has been first studied by \cite{obe09} using a broad-band discharge lamp with a spectral emission peaking at 10.2 eV (Lyman-$\alpha$). A photodesorption rate of methanol of $\sim$ 10$^{-3}$ molecules/photon was derived, and is now largely used in astrochemical models \citep{guz13,wal14}. In the mean time more recent studies have extended our knowledge of UV photodesorption process. The critical role of the ice composition has been highlighted in the case of CO:N$_2$ and CO:CO$_2$ binary ices \citep{ber13,fil14}, in which the absorption of UV photons by condensed CO offers a new desorption pathway below 10 eV for the co-adsorbed molecules. In addition, the photodissociation of molecules plays an important part in the overall desorption process. In particular, the exothermicity of recombination reactions of photofragments in the condensed phase can lead to the desorption of the product, as initially proposed by \cite{wil68} and verified since then for H$_2$O, O$_2$ and CO$_2$ \citep{yab13,fay13,fil14,mar15}. Since both composition of the ices and photodissociation are involved, it is likely that the photodesorption of methanol ice is a rather complex process, and that a proper estimation of the photodesorption rate of the intact methanol from methanol-containing ices requires also the evaluation of the formation and desorption of the photofragments. 

In this work, we have studied the UV photodesorption of the CH$_3$OH from pure methanol ice, but also from ice mixtures of CH$_3$OH with CO with controlled stoichiometry. Data are recorded as a function of the irradiating photon energy, which allows for getting insight in the involved molecular mechanims, and for deriving absolute photodesorption rates applicable to any UV field. The aim is here to re-estimate the photodesorption rate of the methanol ice for astronomically relevant temperatures, and to access the desorption rates of photofragments, that is dissociation products of the condensed CH$_3$OH molecules by UV photons. In section 2, the experimental procedures are introduced. The results and their astrophysical implications are presented and discussed in section 3.

\section{Methods}

The studies are realized in the ‘SPICES’ (Surface Processes \& ICES) setup of the UPMC (Universit\'e Pierre \& Marie Curie), under ultrahigh vacuum (UHV) conditions ($P \sim 1\times10^{-10}$ Torr). The substrate on which ices are grown is a polycrystalline gold surface. It is mounted on the tip of a rotatable cold head that can be cooled down to $\sim$ 9 K by means of a closed cycle helium cryostat. The temperature remains stable within 0.1 K even under UV irradiation. The molecular ices are grown by exposing the cold sample to a partial pressure of methanol gas or gaseous mixture of methanol and CO. This is realized by a tube positioned a few millimeters away from the surface, which allows the ice growth without a substantial increase of the base pressure in the chamber (less than a few $10^{-10}$ Torr). Gaseous mixtures of methanol (99,9 \% purity, Sigma-Aldricht) and CO (99 \% purity, Alphagaz) can be prepared before entering the chamber, with CH$_3$OH:CO ratios varying between 1:1 and 1:100. The control of the composition is provided by recording the partial pressure of the two gases by capacitive gauges in the gas dosing system. The exact ices thickness and composition is further calibrated by the temperature programmed desorption (TPD) technique \citep{dor15}. The growing protocol allows for deposition of ice with a precision better than 1 ML (one monolayer ML being equivalent to $\sim 10^{15}$ molecules.cm$^{-2}$).

The irradiation of the ices has been realized at the SOLEIL synchrotron facility, on the DESIRS beamline. Monochromatic photons in the energy range of 7 - 14 eV have been used, thanks to a windowless coupling of the UHV chamber on the beamline. The DESIRS beamline \citep{nah12} provides VUV beams in two working regimes: high flux, low resolution ($\sim$ 0.5 eV) output of an undulator, or lower flux, higher resolution ($\sim$ 25 meV) output provided by a grating monochromator, and for which the photon energy can be continuously tuned. In both cases, the higher harmonics of the ondulator output are suppressed by a rare gas filter. Due to the low efficiency of the methanol ice photodesorption, only the high flux regime has been used in this study. Typical photon fluxes are measured using both a gold mesh and a photodiode, and varied from 0.5 to 1.5 $\times 10^{15}$ photons.s$^{-1}$.cm$^{-2}$ depending on the photon energy. 

The photodesorption is studied by recording in the gas phase the signal of desorbed species by means of a quadripolar mass spectrometer (QMS), while irradiating the ice with photons of several fixed energies in the 7 - 14 eV range. This results in a photodesorption spectrum as a function of the photon energy. In practice, the irradiation time for each energy is kept below 5 s in order to limit the photo-aging of the ice, but this is still much higher than the acquisition time of the QMS (100 ms). It has been checked that the irradiations performed several times on the same ice led to the same photodesorption spectra within the experimental error bars, therefore ruling out a non-negligible aging effect on the extracted photodesorption rates. Infrared spectroscopy, which is used throughout the irradiation steps, confirms that our irradiation conditions lead to only slight chemical alterations of the ice. 

The QMS signal, corrected from contributions from the background residual gas, can be calibrated to photodesorption rate, expressed in desorbed molecules per incident photon. Each gas phase species is probed by monitoring the mass signal of its corresponding intact cation: CH$_3$OH, H$_3$CO, H$_2$CO, CO, OH and CH$_3$ are recorded by selecting mass 32, 31, 30, 28, 17 and 15 amu, respectively. On each mass channel, the contributions of the fragment ions from dissociative electron-impact ionization into the QMS of heavier species like CH$_3$OH, H$_3$CO and H$_2$CO \citep{sri96,vac09} have been evaluated and subtracted from the final signal. Concerning the fragment H$_3$CO (mass 31 amu), mass spectrometry is unable to differentiate between the methoxy radical (CH$_3$O) and the hydroxymethyl radical (CH$_2$OH).

The calibration procedure follows three steps: (i) for a given mass, the signal is corrected by the incident photon flux at each photon energy; (ii) using the well-known photodesorption spectrum of CO from a pure CO ice \citep{fay11} a proportionality coefficient between flux-corrected signal and the photodesorption rate is extracted, which takes into account the $\sim$ 0.5 eV width of the photon energy profile; (iii) for a given species, the resulting photodesorption rate is corrected to take into account the apparatus function of the QMS for different masses and the difference of non-dissociative electron-impact ionization cross section with the one for CO$^+$ from CO. These values are available in the literature for CH$_3$ \citep{bai84}, CO \citep{fre90}, CH$_3$OH \citep{sri96}, H$_2$CO \citep{vac09} and OH \citep{jos01}. For CH$_3$O/CH$_2$OH literature data are not vailable, and we have considered a non-dissociative electron impact ionization cross section of 1.25 \AA$^2$, which is halfway between the values for CH$_3$OH (1.17 \AA$^2$) and for H$_2$CO (1.3 \AA$^2$).

\section{Results and discussion}

\subsection{Photodesorption from pure CH$_3$OH ices} 

Figure 1 presents photodesorption rates of CH$_3$OH (Fig. 1a) and fragments of CH$_3$OH (Fig. 1b) as a function of the incident photon energy from a pure 20 ML thick CH$_3$OH ice at 10 K. The absorption spectrum of a pure CH$_3$OH ice from \cite{cru14} is given for comparison in the 7 - 10.5 eV range. In this study, the gas phase signal of intact methanol is clearly observed above the noise threshold: this is to our knowledge the first direct detection of photodesorbing intact methanol molecules from physisorbed systems. For the desorption of intact methanol and of its photofragments, the energy-dependent photodesorption rates follow the absorption spectrum of the CH$_3$OH ice in the low energy range, with a monotonic increase between 7 and 10 eV after which it stabilizes. This gives further confidence that the observed photodesorption originates from the absorption of VUV photons in the ice, ruling out substrate-mediated processes. 

The photodesorption rate of intact CH$_3$OH from pure CH$_3$OH ice (Fig. 1a) is found to be $\sim$ 10$^{-5}$ molecules/photon. This value is two orders of magnitude lower than what was previously evaluated in the study of \cite{obe09}, in which an average rate of $\sim$ 10$^{-3}$ molecules/photon was derived. However, in the work of \cite{obe09}, only the loss of condensed methanol was probed, and then correlated by the authors to the intact methanol desorption. In our study, the gas phase detection of the photodesorbing species reveals that not only the intact methanol is observed. Indeed, a series of photofragments is released into the gas during irradiation (Fig. 1b). The overall photodesorption process is dominated by the desorption of CO, with a rate of $\sim$ 10$^{-4}$ molecules/photon (the CO datapoints intensity is divided by a factor of 6 to compare all individual photofragments in one plot -- Fig. 1b). Other photofragments, H$_2$CO, CH$_3$ and OH, have an efficiency comparable to the photodesorption of intact CH$_3$OH ($\sim$ 10$^{-5}$ molecules/photon range).  Finally, a very weak signal associated with the CH$_3$O or CH$_2$OH radical can be seen, but, as the signal falls within our noise level, only an upper limit for the CH$_3$O/CH$_2$OH photodesorption rate of $\sim 3\times10^{-6}$ molecules/photon can be confidently given. The desorption of the intact molecule is relatively a weak channel of the overall desorption process. This supposes that the rate extracted in \cite{obe09} from the condensed phase CH$_{3}$OH disapperance takes into account the desorption of all the photofragments, mainly CO, and not only of the intact methanol. This gives us confidence that the value given by \cite{obe09}, assigned solely to the desorption of intact methanol, is largely overestimated.

\subsection{Photodesorption from CO:CH$_3$OH mixed ices}

The photodesorption has been studied on methanol embedded in CO-rich ices, with CO:CH$_3$OH mixing ratios ranging from 1:4 to 1:50. In each ice, the total amount of CH$_3$OH molecules is kept constant and equivalent to 20 ML. The choice of CO as co-adsorbate is motivated by (i) the fact that methanol is believed to be formed and mixed in CO-rich ices \citep{wat02,fuc09,her09,cup11} and (ii) the already known efficient energy transfer from excited CO to co-adsorbed N$_2$ or CO$_2$ that triggers their desorption with the condensed CO excitation pattern \citep{ber13,fil14}. Fig. 2 presents the photodesorption spectra of CH$_3$, OH, H$_2$CO, CH$_3$O/CH$_2$OH and CH$_3$OH obtained from several condensed mixtures of CH$_3$OH and CO at 10 K. 

It is seen that the desorption of the fragments CH$_3$, OH and H$_2$CO matches the photodesorption spectra from the pure methanol ice. In this case, these species are expected to originate from the direct dissociation of intact condensed methanol following:
\begin{equation}
\mathrm{CH_3OH_{solid} + h\nu \longrightarrow CH_{3gas} + OH_g}
\end{equation} 
\begin{equation}
\mathrm{CH_3OH_{solid} + h\nu \longrightarrow H_2CO_{gas} + H_2 \ (or \  2H)}
\end{equation} 
during which excess kinetic energy allows the fragments to overcome the adsorption barriers. Their photodesorption rates do not depend on the CO concentration in the ices, which rules out subsequent condensed phase chemistry which should vary with increasing dilution. This is further supported by the correlation between photodesorption rates of OH and CH$_3$, found identical within the error bars, indicating that they both originate from the same dissociation event of methanol. The mass signal of H$_2$ has not been monitored during the experiments. Finally, the CO photodesorption from the mixture mainly corresponds to the photodesorption from pure CO ices, which is much larger than the one from the pure methanol ice. Thus, the study of the CO:methanol mixtures does not give hints on the CO production mechanism in the case of pure methanol ices, which may originate from several steps of dehydrogenation of the CH$_{3}$OH condensed molecule.

The photodesorption of intact CH$_3$OH and CH$_3$O/CH$_2$OH follows a different trend with the increasing methanol dilution in the ice. The CH$_3$OH desorption is only observed in the case of pure methanol ice. When the methanol is mixed with CO (ratio higher than 1/4), the CH$_3$OH mass signal drops below our detection limit, and therefore only a higher limit for the photodesorption rate of ~ 3.10$^{-6}$ molecules/photon can be given. Interestingly, no intact CH$_3$OH desorption is found upon the excitation of the CO matrix, ruling out an indirect DIET mechanism similar to what has been highlighted in the case of CO:N$_2$ or CO:CO$_2$ icy mixtures \citep{ber13,fil14}. Moreover, the desorption of CH$_3$O/CH$_2$OH fragments, which is only barely detected in the case of pure methanol ice, is clearly visible from mixed CO:methanol ice with a CH$_3$OH ratio below 1/4. This anti-correlation between CH$_3$OH and H$_3$CO desorption as a function of CO concentration in the ice suggests that the desorption of intact CH$_3$OH from pure methanol ice originates from an exothermic chemical reaction involving the radical H$_3$CO and leading to methanol reformation. Such a recombination process has already been identified as the origin of the photodesorption of other dissociating molecules such as O$_2$, CO$_2$ or H$_2$O \citep{fay13,fil14,yab13,mar15}. This process becomes inoperative in the case of CO-rich ices, for which the H$_3$CO fragments are free to desorb after their formation. A possible explanation for this finding is that cage effects in CO and CH$_3$OH matrices are different; the CH$_3$OH surrounding of the dissociating molecules in the case of pure methanol ices may favor the recombination of the H$_3$CO fragment after its formation. It has to be noted that the role of condensed phase recombination chemistry for the desorption of methanol was already proposed by \cite{obe09}, although the responsible radical was not identified. 

\subsection{Astrophysical implications}

The energy-dependent photodesorption spectra as presented in Fig. 1 and Fig. 2 allow to derive integrated photodesorption rates for several typical UV fields relevant to different regions of the interstellar medium, that are the UV interstellar radiation field (ISRF) from \cite{mat83}, the dense cloud UV field from \cite{gre87} and the disk UV field from \cite{joh07}. The resulting photodesorption values extracted from pure methanol ice and mixed CO:methanol ices on the basis of the experiments described here are shown in Table 1. The rates for UV fields in dense cores and disks are found identical within the error bars, and very close to the values found for Lyman-$\alpha$ in our photodesorption spectra. The rates established with the ISRF are somewhat different and slightly lower because of the higher contribution of the lower energy photons ($<$ 8 eV) in the UV field, in ranges where the photodesorption is less efficient.

As stated in the introduction, the presence of CH$_3$OH in the gas phase of cold interstellar media, such as prestellar cores and low-UV-field illumination PDRs \citep{guz13,vas14}, is usually interpreted as a result of the photodesorption of condensed methanol. In other cases, such as in protoplanetary disks in which gas phase methanol is still not detected, the photodesorption rates are used to predict the gas phase methanol densities \citep{wal14}. In both cases, models aiming at simulating the chemistry of these regions usually consider photodesorption rates of methanol of ~ 10$^{-3}$ -- 10$^{-4}$ molecules/photon \citep{guz13,wal14}. The present study shows that this value is largely overestimated. In the case of pure methanol ice, the photodesorption rate of intact methanol has been found to lie in the 10$^{-5}$ molecules/photons range, which is one to two orders of magnitude lower than what is generally assumed. Moreover, this rate drops at least below 3.10$^{-6}$ molecules/photons in the case of CH$_3$OH mixed with CO, which is due to the fact that the desorbed methanol presumably originates from a recombination reaction in the ice that involves the formation of the CH$_3$O/CH$_2$OH radicals. The rates extracted for CO:CH$_3$OH mixed ices are particularly interesting since both species are expected to be mixed in the colder interstellar ices \citep{cup11}. No other mixed ice has been tested during our work, but we expect a similar effect since condensed phase chemistry is at stake. This needs however to be confirmed by studies of methanol:H$_2$O mixtures. This supposes that the photodesorption of methanol from mixed, more realistic interstellar ices is less efficient than the desorption from pure methanol ices, and that the rates experimentally extracted from pure ices should be considered as upper limits. Such a significant change in the photodesorption rates for methanol is expected to bring important changes to the abundance of gaseous CH$_3$OH, or even to the snowline location, as predicted from chemical models including gas-grain exchanges. Obviously this raises the question whether the photodesorption of intact methanol as re-evaluated here is really responsible for the observation of gas phase methanol in cold interstellar medium. 

Another important finding of this study is that the photoprocessing of a pure or mixed methanol ice induces the desorption of photofragments, with efficiencies at least as high as the photodesorption of the methanol. This is an important point since the release into the gas phase of radicals such as CH$_3$, OH or CH$_3$O/CH$_2$OH may open new gas phase chemical networks. It therefore appears interesting to re-evaluate whether gas phase chemistry can account or not to the reformation of the methanol by considering the enrichment of the gas phase by the photodesorbing radical from methanol-containing ices. This of course will depend on the efficiency of the involved gas phase reactions. Being conclusive on this question requires quantifying the effects of a drastically decreased photodesorption rate of intact methanol and implementing photodesorption rates of radicals in the astrochemical models. 

Finally, among the desorbing photofragments, the case of the CH$_3$O/CH$_2$OH is particularly interesting. Due to the lack of spectroscopic data, observations of the CH$_2$OH radical have so far never been performed. But, the methoxy CH$_3$O radical has been detected in dark clouds together with more complex organics \citep{cer12}. This radical could participate in the formation of more complex molecules, such as methyl formate HCOOCH$_3$, by gas phase chemical pathways, but its origin is still unclear. Here we show that the UV irradiation of a CO-rich mixed ice comprising methanol can release this radical into the gas  with efficiencies of about 10$^{-5}$ molecules/photon, and therefore propose a possible origin of the methoxy radical as a photodesorbed product of the condensed methanol. Observational mapping of the gas phase CH$_3$O abundances in cold media and search for correlations with gas phase methanol density, grain density and photon flux are needed to test this hypothesis.

\bigskip
\bigskip
\bigskip

We acknowledge SOLEIL for provision of synchrotron radiation facilities under the project 20140100 and we would like to thank Nelson De Oliveira and Gustavo Garcia for assistance on the beamline DESIRS. Financial support from the French CNRS national program PCMI (Physique et Chimie du Milieu Interstellaire), the UPMC labex MiChem, the CNRS PICS program and NWO support from a VICI grant are gratefully acknowledged. We also warmly aknowledge fruitful discussions with C. Walsh, V. Taquet and M. Gerin.

\clearpage

\begin{figure}
\epsscale{.80}
\plotone{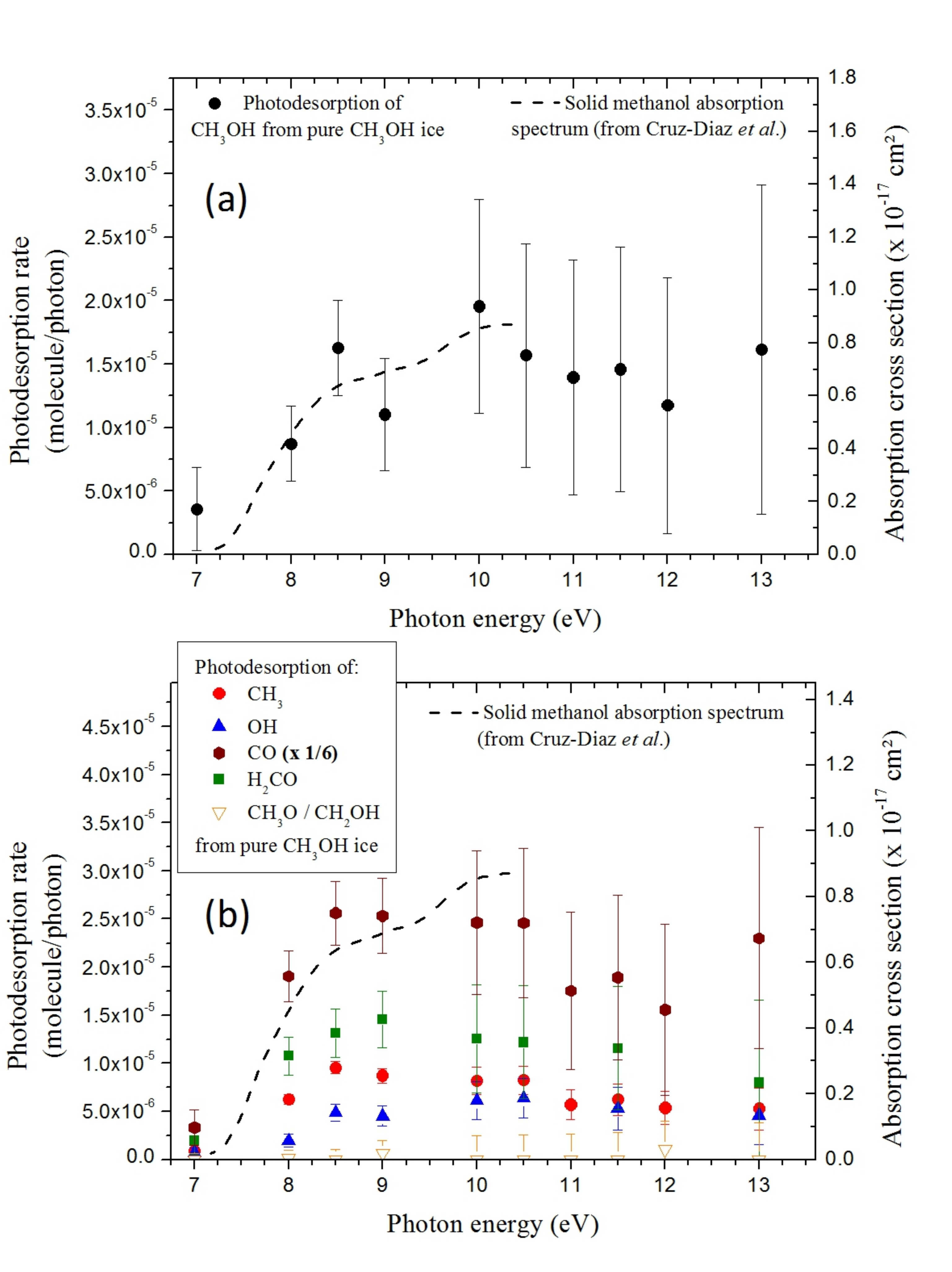}
\caption{(a) Energy-resolved photodesorption rates of the CH$_3$OH molecule from pure 20 ML CH$_3$OH ice condensed on polycrystalline gold at 10 K. (b) Energy-resolved photodesorption rates of photodissociation products of methanol: CO, CH$_3$, OH, H$_2$CO and H$_3$CO, from pure 20 ML CH$_3$OH ice condensed on polycrystalline gold at 10 K. The photodesorption spectrum of CO has been divided by 6 for clarity. In both figures, the absorption spectrum of pure methanol ice, as adapted from \cite{cru14}, is plotted as a dashed line.}.\label{fig1}
\end{figure}

\clearpage

\begin{figure}
\epsscale{1.2}
\plotone{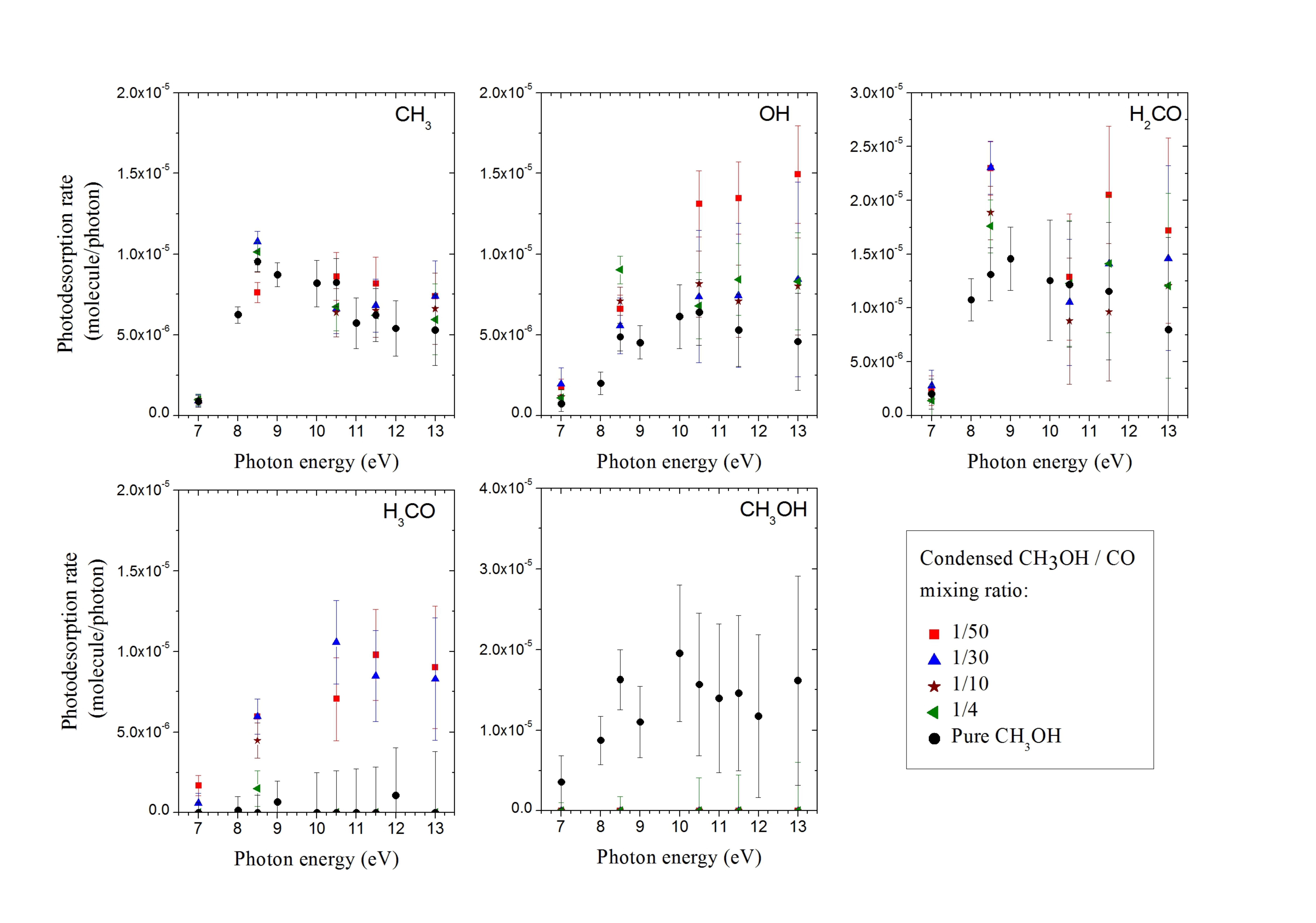}
\caption{Energy-resolved photodesorption rates of CH$_3$, OH, H$_2$CO, H$_3$CO and CH$_3$OH from pure CH$_3$OH and mixed CO:CH$_3$OH ices deposited on polycrystalline gold at 10 K. In each condensed mixing, the total amount of CH$_3$OH is kept constant at 20 ML.}.\label{fig2}
\end{figure}

\clearpage

\begin{table}
\begin{center}
\caption{Integrated photodesorption rates of CH$_3$OH and photofragments of CH$_3$OH from pure CH$_3$OH ice and CH$_3$OH:CO mixtures at 10 K for different astronomical environments. Rates have been derived considering our energy-resolved photodesorption rates shown in Fig. 1 and Fig. 2 and several interstellar-relevant UV fields, between 7 and 14 eV.}
\begin{tabular}{cccc}
\hline
\hline
\multirow{4}{*}{Photodesorbed species} & \multirow{4}{*}{CH$_3$OH ice} & \multicolumn{2}{c}{Integrated photodesorption rate }\\
& & \multicolumn{2}{c}{($\times$ 10$^{-5}$ molecule/photon)} \\
& & \multirow{2}{*}{ISRF $^a$} & Prestellar cores $^b$  \\
& &  & and protoplanetary disks $^c$  \\
\hline
\multirow{2}{*}{CH$_3$OH} & Pure & 1.2 $\pm$ 0.6 & 1.5 $\pm$ 0.6 \\
& Mixed with CO & \multicolumn{2}{c}{$<$ 0.3} \\
\multirow{2}{*}{CH$_3$O / CH$_2$OH} & Pure &  \multicolumn{2}{c}{$<$ 0.3} \\
& Mixed with CO & 0.7 $\pm$ 0.3 & 0.8 $\pm$ 0.5  \\
CO & Pure & 19 $\pm$ 3 & 21 $\pm$ 3 \\
H$_2$CO & Pure \& Mixed with CO & 0.7 $\pm$ 0.3 & 1.2 $\pm$ 0.4 \\
OH & Pure \& Mixed with CO & 0.3 $\pm$ 0.1 & 0.7 $\pm$ 0.3 \\
CH$_3$ & Pure \& Mixed with CO & 0.3 $\pm$ 0.1 & 0.8 $\pm$ 0.4 \\
\hline
\end{tabular}
\tablecomments{Using UV field from $^a$\cite{mat83}, $^b$\cite{gre87}, $^c$\cite{joh07}.}
\end{center}
\end{table}

\end{document}